\documentclass[runningheads]{svmult}
\usepackage{makeidx}   
\usepackage{graphicx}  
\usepackage{subeqnar}  
\usepackage{multicol}  
\usepackage{physprbb}  
\makeindex             
\begin{document}
\title*
{Experimental Results on Two-Photon Physics from LEP
\thanks{Invited talk given at the Ringberg Workshop 
{\it New Trends in HERA Physics 1999},  
May 30 $-$ June 4, 1999, to appear in the Proceedings.}
}
\toctitle{Experimental Results on Two-Photon Physics from LEP}
%
\titlerunning{Experimental Results on Two-Photon Physics from LEP}
%
\author{Richard Nisius}
\authorrunning{Richard Nisius}
%
%
\institute{CERN, CH-1211 Gen\`eve 23, Switzerland}
\maketitle              
\begin{abstract}
 This review covers selected results from the LEP experiments on the 
 structure of quasi-real and virtual photons.
 The topics discussed are
 the total hadronic cross-section for photon-photon scattering,
 hadron production,
 jet cross-sections,
 heavy quark production for photon-photon scattering,
 photon structure functions,
 and cross-sections for the exchange of two virtual photons.
\end{abstract}
%
%
\section{INTRODUCTION}
\label{sec:intro}
\begin{figure}[htb]
\begin{center}
{\includegraphics[width=0.6\linewidth]{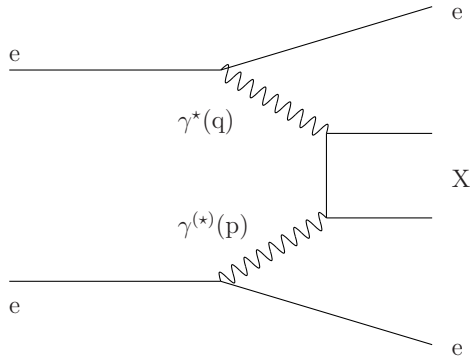}}
\end{center}
\caption{
         A diagram of the reaction 
         ${\mathrm e}{\mathrm e}\rightarrow {\mathrm e}{\mathrm e} X$,
         proceeding via the exchange of two photons.
        }\label{fig:fig01}
\end{figure}

 The photon structure has been investigated in detail at LEP based 
 on the scattering of two electrons\footnote{Fermions and anti-fermions 
 are not distinguished, for example, electrons and positrons are referred 
 to as electrons. The natural system of units, which means, 
 $c=\hbar=1$ is used.} proceeding via the
 exchange of two photons, as shown in Figure~\ref{fig:fig01}.
%
%
 The reactions are classified depending on the virtualities of the photons,
 with $-q^2=Q^2$ and $-p^2=P^2$, and on the nature of the final state $X$.
 If a photon has small virtuality, and the corresponding electron is not 
 observed in the experiment, it is called a quasi-real photon,
 $\gamma$.
 If the electron is observed, the photon is far off-shell, 
 and the virtual photon is denoted with $\gamma^\star$.
 As the two photons can either be quasi-real or virtual the 
 reactions are classified as 
 $\gamma\gamma$ scattering (photon-photon scattering or anti-tagged events), 
 $\gamma\gamma^\star$ scattering 
 (deep inelastic electron-photon scattering or single tag events),
 and $\gamma^\star\gamma^\star$ scattering (double tag events).
 In each of the classes, different aspects of the photon structure 
 can be investigated.
 Due to space limitations not all results can be reviewed here,
 only a personal selection has been chosen driven by the relevance
 of the different topics in the context of this workshop on HERA physics.
 The main results not covered here concern resonance production
 and glueball searches, which are described in 
 Ref.~\cite{L3C-9501L3C-9702L3C-9802L3C-9901L3C-9904OPALPR248}.
%
%
\section{RESULTS FROM $\gamma\gamma$ SCATTERING}
\label{sec:gg}
 The $\gamma\gamma$ scattering reaction has the largest hadronic
 cross-section at LEP2 energies.
 The main topics studied are the  total hadronic cross-section for 
 photon-photon scattering, $\sigma_{\gamma\gamma}$,
 and more exclusively, hadron production, jet cross-sec\-tions 
 and the production of heavy quarks.
%
%
\subsection{THE TOTAL PHOTON-PHOTON CROSS-SECTION}
\label{sec:ggtot}
 The measurement of $\sigma_{\gamma\gamma}$ is both,
 interesting and challenging. 
 It is interesting, because in the framework of Regge theory 
 $\sigma_{\gamma\gamma}$ can be
 related to the total hadronic cross-sections for photon-proton and 
 hadron-hadron scattering, $\sigma_{\gamma{\rm p}}$ and 
 $\sigma_{{\rm h}{\rm h}}$, and a slow rise with the photon-photon 
 center-of-mass energy, $s=W^2$, is predicted.
 It is challenging, firstly, because experimentally
 the determination of the hadronic invariant mass, $W$, is very difficult
 due to limited acceptance and resolution for the hadrons created in the 
 reaction and secondly, because the composition of different event classes, 
 for example, diffractive and quasi-elastic processes, is rather uncertain,
 which affects the overall acceptance of the events.
 The first problem is dealt with by determining $W$ from the 
 visible hadronic invariant mass using unfolding programs.
 The second uncertainty is taken into account by using two models,
 namely 
 PHOJET~\cite{ENG-9501ENG-9601}
 and PYTHIA~\cite{SJO-9401}, for 
 the description of the hadronic final state and for the correction from
 the accepted cross-section to $\sigma_{\gamma\gamma}$, leading to 
 the largest uncertainty of the result.
 \par
%
\begin{figure}[htb]
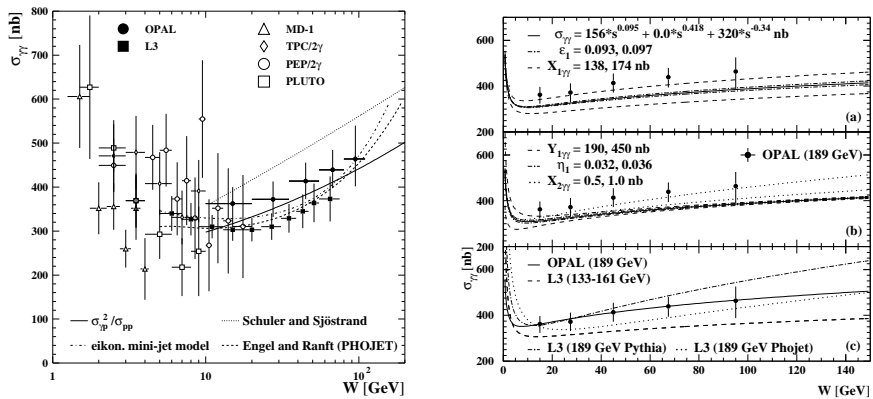

\begin{center}
{\includegraphics[width=0.48\linewidth]{./nisius.2}}
{\includegraphics[width=0.50\linewidth]{./nisius.3}}
\end{center}
\caption{
         The published results on $\sigma_{\gamma\gamma}$ as a function
         of $W$ (right), and an illustration of the spread of the 
         fit results to various data (left).
        }\label{fig:fig02}
\end{figure}
%
 The published measurements of $\sigma_{\gamma\gamma}$
 by L3~\cite{L3C-9704} and by OPAL~\cite{OPALPR278} are shown in
 Figure~\ref{fig:fig02}(left).
 Both results show a clear rise as a function of $W$.
 The cross-section  $\sigma_{\gamma\gamma}$ is interpreted within the
 framework of Regge theory, motivated by the fact that 
 $\sigma_{\gamma{\rm p}}$ and $\sigma_{{\rm h}{\rm h}}$ are well described
 by Regge parametrisations using terms to account for
 pomeron and reggeon exchanges.
 The originally proposed form of the Regge parametrisation for 
 $\sigma_{\gamma\gamma}$ is
%
 \begin{equation}
 \sigma_{\gamma\gamma}(s) =
 X_{1\gamma\gamma} s^{\epsilon_1}+ Y_{1\gamma\gamma} s^{-\eta_1}\, ,
 \end{equation}
%
 where $s$ is taken in units of GeV$^2$.
 The first term in the equation is due to soft pomeron exchange
 and the second term is due to reggeon exchange.
 The exponents $\epsilon_1$ and $\eta_1$ are assumed to be universal.
 The presently used values of $\epsilon_1=0.095\pm0.002$ and 
 $\eta_1=0.034\pm0.02$ are taken from Ref.~\cite{PDG-9801}.
 The parameters were obtained by a fit to the total hadronic cross-sections
 of pp, ${\rm p}\bar{{\rm p}}$, $\pi^{\pm}$p, K$^{\pm}$p, 
 $\gamma$p and $\gamma\gamma$ scattering reactions.
 The coefficients $X_{1\gamma\gamma}$ and $Y_{1\gamma\gamma}$ have to 
 be extracted from the $\gamma\gamma$ data.
 The values obtained in Ref.~\cite{PDG-9801} by a fit to older 
 $\gamma\gamma$ data, including those of L3 from Ref.~\cite{L3C-9704},
 are $X_{1\gamma\gamma}=(156 \pm 18)$~nb and
 $Y_{1\gamma\gamma}=(320 \pm 130)$~nb.
 Recently an additional hard pomeron component has been suggested in 
 Ref.~\cite{DON-9801} leading to
%
 \begin{equation}
 \sigma_{\gamma\gamma}(s)= X_{1\gamma\gamma} s^{\epsilon_1}+
 X_{2\gamma\gamma} s^{\epsilon_2}+Y_{1\gamma\gamma} s^{-\eta_1}\, ,
 \label{eqn:sigma}
 \end{equation}
%
 with a proposed value of $\epsilon_2=0.418$ and an expected
 uncertainty of $\epsilon_2$ of about $\pm 0.05$.
%
\begin{figure}[htb]
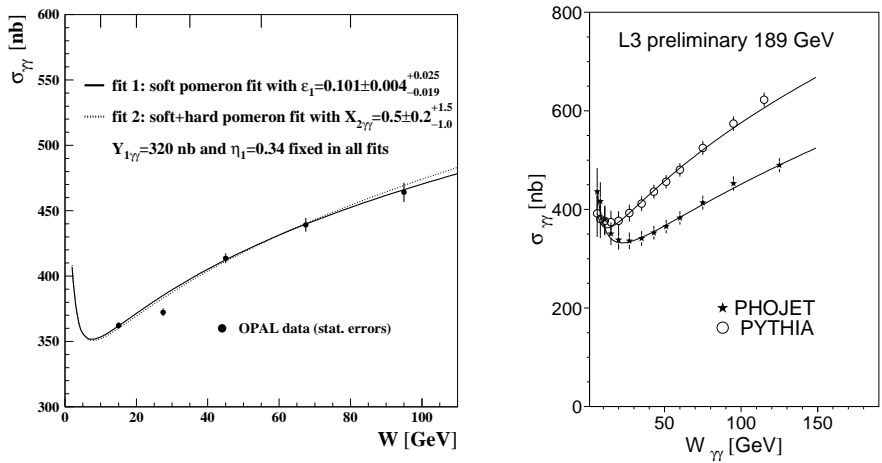

\begin{center}
{\includegraphics[width=0.55\linewidth]{./nisius.4}}
{\includegraphics[width=0.41\linewidth]{./nisius.5}}
\end{center}
\caption{
         Fits to the total hadronic cross-section 
         for photon-photon scattering as a function of $W$
         for OPAL data at $\sqrt{s}_{\rm ee}=161-189$~GeV (left), 
         and for L3 data at $\sqrt{s}_{\rm ee}=189$~GeV using two different
         Monte Carlo models for correcting the data (right).
         }\label{fig:fig03}
\end{figure}
%
 Different fits to the data have been performed by the experiments.
 \par
 The interpretation of the results is very difficult, because, firstly 
 the parameters are highly correlated, secondly, the main region
 of sensitivity to the reggeon term is not covered by the OPAL measurement
 and thirdly, different assumptions have been made when performing
 the fits.
 The correlation of the parameters of Eq.~(\ref{eqn:sigma}) can be 
 clearly seen in Figure~\ref{fig:fig02}(right a,b), where the theoretical
 predictions are shown, exploring the uncertainties 
 for the soft pomeron term in (a) and for the reggeon as well as for
 the hard pomeron term in (b), using the central values and
 errors quoted in Ref.~\cite{PDG-9801}.
 It is clear from Figure~\ref{fig:fig02}(right a,b) that by
 changing different parameters in (a) and (b) a very similar 
 effect on the rise of the total-cross section can be achieved.
 Figure~\ref{fig:fig02}(right c) shows the spread of the best fit curves 
 for various data and various fit assumptions explained  below.
 In Figure~\ref{fig:fig02}(right a-c) in addition the results from 
 Ref.~\cite{OPALPR278} are shown
 to illustrate the size of the experimental uncertainties.
 \par
 Examples of different fits are shown in Figure~\ref{fig:fig03} 
 taken from Refs.~\cite{OPALPR278,CSI-9901}.
 They yield the following results:
 \par
 $\bullet$
 The OPAL data, within the present range of $W$, can be accounted for without
 the presence of the hard pomeron term. When fixing all exponents
 and $Y_{1\gamma\gamma}$ to the values listed above the fit yields
 $X_{2\gamma\gamma}=(0.5\pm0.2^{+1.5}_{-1.0})$~nb, which is not significantly
 different from zero, and $X_{1\gamma\gamma}=(182 \pm 3 \pm 22)$~nb, which is 
 consistent with the values from Ref.~\cite{PDG-9801}.
 Using $X_{2\gamma\gamma}=0$ and leaving only $\epsilon_1$ and 
 $X_{1\gamma\gamma}$ as free parameters results in
 $\epsilon_1=0.101\pm0.004^{+0.025}_{-0.019}$ and
 $X_{1\gamma\gamma}=(180 \pm 5^{+30}_{-32})$~nb, 
 Figure~\ref{fig:fig02}(right, c, full), again consistent with
 Ref.~\cite{PDG-9801}.
 \par
 $\bullet$
 In all fits performed by L3 the hard pomeron term is set to zero.
 The L3 data from Ref.~\cite{L3C-9704}
 can be fitted using the old values for the exponents of
 $\epsilon_1=0.0790\pm0.0011$ and $\eta_1=0.4678\pm0.0059$ 
 from  Ref.~\cite{PDG-9601}
 leading to $X_{1\gamma\gamma}=(173 \pm 7)$~nb and
 $Y_{1\gamma\gamma}=(519 \pm 125)$~nb, Figure~\ref{fig:fig02}(right, c, dash).
 The L3 data at $\sqrt{s}_{\rm ee}=189$~GeV indicate a faster rise with
 energy.
 Using $\epsilon_1=0.95$ and $\eta_1=0.34$, and the PHOJET Monte Carlo
 for correcting the data, leads to
 $X_{1\gamma\gamma}=(172\pm 3)$~nb and $Y_{1\gamma\gamma}=(325 \pm 65)$~nb, 
 but the confidence level of the fit is only 0.000034~\cite{CSI-9901}.
 Fixing only the reggeon exponent to $\eta_1=0.34$ leads to 
 $\epsilon_1 = 0.222 \pm 0.019 / 0.206 \pm 0.013$, 
 $X_{1\gamma\gamma}=(50   \pm   9)\, /\, (78  \pm  10)$~nb and 
 $Y_{1\gamma\gamma}=(1153 \pm 114)\, /\, (753 \pm 116)$~nb, when using
 PHOJET/PYTHIA, Figure~\ref{fig:fig02}(right, c, dot/dot-dash).
 \par
 In summary, the situation is unclear at the moment with OPAL being 
 consistent with the universal Regge prediction, whereas L3
 indicating a faster rise with $s$ for the data at 
 $\sqrt{s}_{\rm ee}=189$~GeV. 
 In addition, the L3 data taken at different center-of-mass energies
 show a different behaviour of the measured cross-section,
 with the data taken at $\sqrt{s}_{\rm ee}=133-161$~GeV being lower, 
 especially for $W<30$~GeV.
%
%
\subsection{THE PRODUCTION OF CHARGED HADRONS}
\label{sec:gghad}
\begin{figure}[htb]
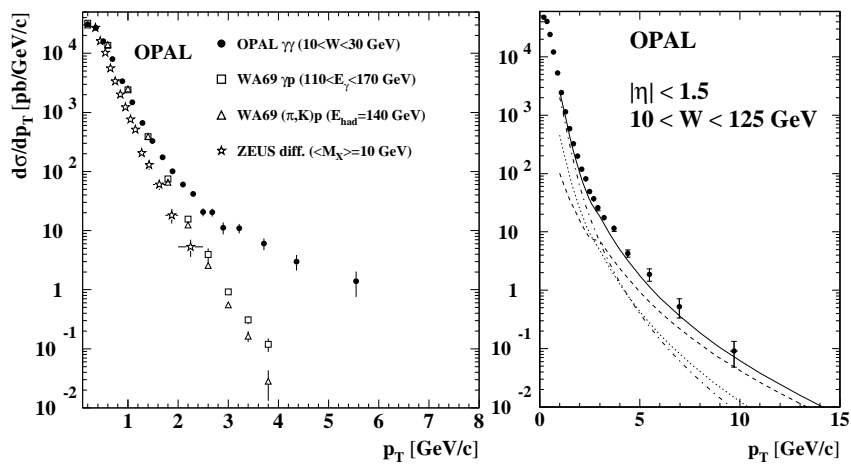

\begin{center}
{\includegraphics[width=0.515\linewidth]{./nisius.6}}
{\includegraphics[width=0.390\linewidth]{./nisius.7}}
\end{center}
\caption{
         The differential single particle inclusive cross-section
         for hadron production in photon-photon scattering
         at $\sqrt{s}_{\rm ee}=161-172$~GeV
         compared to other experiments for $10<W<30$~GeV (left),
         and compared to next-to-leading order calculations for 
         $10<W<125$~GeV (right).
        }\label{fig:fig04}
\end{figure}

 The production of charged hadrons is sensitive to the structure of
 the photon-photon interactions without theoretical and experimental
 problems related to the definition and reconstruction of jets.
 The two main results from the study of hadron production at 
 LEP are shown in Figure~\ref{fig:fig04}.
 \par
%
%
 In Figure~\ref{fig:fig04}(left) the differential single particle inclusive
 cross-section ${\mathrm{d}}\sigma/{\mathrm{d}}p_{\rm T}$ 
 for charged hadrons for $\gamma\gamma$ scattering as obtained by
 OPAL~\cite{OPALPR241}, with $10<W<30$~GeV, is shown, together with
 results from $\gamma {\rm p}$, $\pi {\rm p}$ and $ {\rm K}{\rm p}$ 
 scattering from WA69 with a hadronic invariant mass of $16$~GeV. 
 The WA69 data are normalised to the $\gamma\gamma$ data at 
 $p_{\rm T}\approx 0.2$~GeV. 
 In addition, ZEUS data from Ref.~\cite{ZEU-9502}
 on charged particle production in $\gamma {\rm p}$
 scattering with a diffractively dissociated photon are shown.
 These data have an average invariant mass of the diffractive system of 
 $10$~GeV, and again they are normalised to the OPAL data. 
 In Figure~\ref{fig:fig04}(right) the differential single particle inclusive
 cross-section for $10<W<125$~GeV is compared to next-to-leading order
 QCD predictions. The main findings are:
 \par
 $\bullet$
 The spectrum of transverse momentum of charged hadrons in photon-photon
 scattering is much harder than in the case of photon-proton, hadron-proton
 and `photon-Pomeron' interactions. This can be attributed to the direct
 component of the photon-photon interactions.
 \par
 $\bullet$
 The production of charged hadrons is found to be well described by the
 next-to-leading order QCD predictions from Ref.~\cite{BIN-9601} over a wide
 range of $W$.
 These next-to-leading order calculations are based on the QCD partonic 
 cross-sections, the next-to-leading order GRV parametrisation of the parton 
 distribution functions and on fragmentation functions fitted to 
 e$^+$e$^-$ data.
 The renormalisation and factorisation scales are set equal to $p_{\rm T}$.
%
%
\subsection{JET PRODUCTION}
\label{sec:ggjet}
 Jet production is the classical way to study the partonic structure
 of particle interactions.
 At LEP the di-jet cross-section in $\gamma\gamma$ scattering was studied
 in Ref.~\cite{OPALPR250} at $\sqrt{s}_{\rm ee}=161-172$~GeV using
 the cone jet finding algorithm with $R=1$.
 Three event classes are defined, direct, single-resolved and 
 double-resolved interactions. Here, direct means that the photons as a whole
 take part in the hard interaction, whereas resolved means that a parton of
 a hadronic fluctuation of the photon participates in the hard scattering
 reaction.
 Experimentally, direct and double-resolved interactions can be 
 clearly separated using the quantity
%
 \begin{equation}
 x_\gamma^\pm = \frac{\sum_{\rm jets=1,2}(E \pm p_z)}
                     {\sum_{\rm hadrons}(E \pm p_z)},
 \end{equation}
%
 whereas a selection of single-resolved events cannot be achieved 
 with high purity.
 Ideally, in leading order direct interactions have $x_\gamma^\pm=1$, 
 however, due to resolution and higher order corrections the measured
 values of $x_\gamma^\pm$ are smaller.
 Experimentally, samples containing large fractions of direct events
 can be selected by requiring $x_\gamma^\pm>0.8$, and samples containing 
 large fractions of double-resolved events by using $x_\gamma^\pm<0.8$.
 \par
%
\begin{figure}[htb]
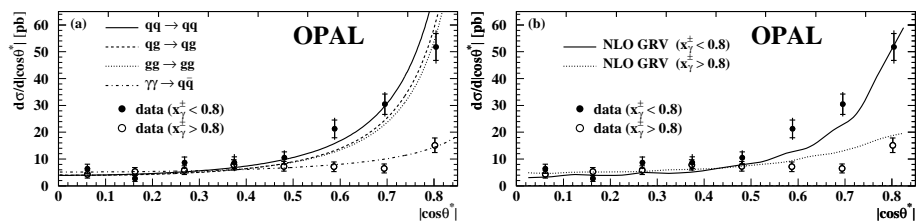

\mbox{}\vspace{-0.7cm}
\begin{center}
{\includegraphics[width=0.49\linewidth]{./nisius.8}}
{\includegraphics[width=0.49\linewidth]{./nisius.9}}
\end{center}
\caption{
         The angular dependence of 
         di-jet production at $\sqrt{s}_{\rm ee}=161-172$~GeV
         compared to leading order matrix elements (left)
         and to next-to-leading order (NLO) predictions (right).
        }\label{fig:fig05}
\mbox{}\vspace{-0.5cm}
\end{figure}
%
 The measurement of the distribution of $\cos\theta^\star$,
 the cosine of the scattering angle in the photon-photon centre-of-mass
 system, allows for a test of the different matrix elements contributing
 to the reaction.
 The scattering angle is calculated from the jet rapidities in the 
 laboratory frame using
%
 \begin{equation}
 \cos\theta^\star = \tanh \frac{\eta^{\rm jet1}-\eta^{\rm jet2}}{2}.
 \end{equation}
%
 In leading order the direct contribution $\gamma\gamma\rightarrow q\bar{q}$
 leads to an angular dependence of the form $(1-\cos^2\theta^\star)^{-1}$,
 whereas double-resolved events, which are dominated by gluon induced
 reactions, are expected to behave approximately as 
 $(1-\cos^2\theta^\star)^{-2}$.
 The steeper angular dependence of the double-resolved interactions 
 can be clearly seen in Figure~\ref{fig:fig05}(left), where the shape
 of the di-jet cross-section, for events with di-jet masses above 
 $12$~GeV and average rapidities of $\vert(\eta_1+\eta_2)/2\vert<1$,
 is compared to leading order predictions.
 In addition, the shape of the angular distribution observed in the data
 is roughly described by the next-to-leading order prediction
 from Refs.~\cite{KLA-9801}, Figure~\ref{fig:fig05}(right).
 In both cases the theoretical predictions are normalised to the data
 in the first three bins.
 \par
 These next-to-leading order calculations well account for the observed
 inclusive differential di-jet cross-section, 
 ${\rm d}\sigma/{\rm d}E^{\rm jet}_{\rm T}$,
 as a function of jet transverse energy, $E^{\rm jet}_{\rm T}$, 
 for di-jet events with pseudorapidities $|\eta^{\rm jet}|<2$.
%
\begin{figure}[htb]
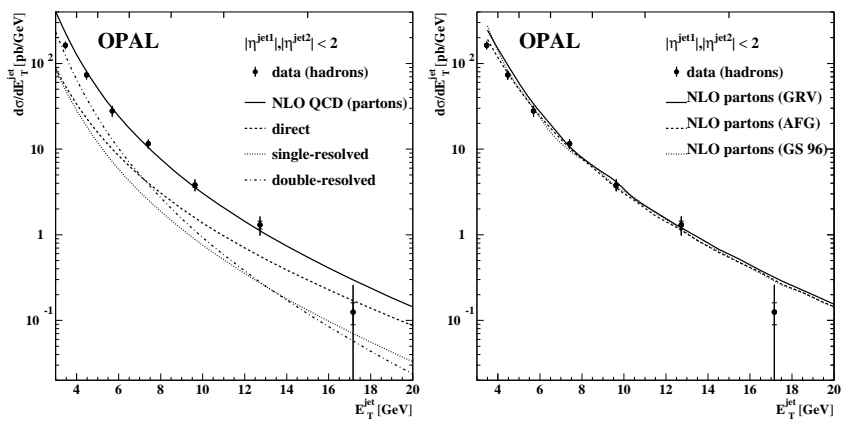

\mbox{}\vspace{-0.7cm}
\begin{center}
{\includegraphics[width=0.45\linewidth]{./nisius.10}}
{\includegraphics[width=0.45\linewidth]{./nisius.11}}
\end{center}
\caption{
         The $E^{\rm jet}_{\rm T}$ dependence of 
         di-jet production at $\sqrt{s}_{\rm ee}=161-172$~GeV
         compared to next-to-leading order (NLO) predictions
         for different event classes (left) and for
         different parametrisations of the parton distribution 
         functions of the photon (right).
        }\label{fig:fig06}
\end{figure}
%
 As expected the direct component can account for most of the cross-section 
 at large $E^{\rm jet}_{\rm T}$, whereas the region of low 
 $E^{\rm jet}_{\rm T}$ is dominated by the double-resolved contribution, 
 shown in Figure~\ref{fig:fig06}(left).
 The calculations from Refs.~\cite{KLApriv} for three different 
 next-to-leading order parametrisations of the parton distribution 
 functions of the photon are in good agreement with
 the data shown in Figure~\ref{fig:fig06}(right), except in the first bin,
 where theoretical as well as experimental uncertainties are large.
 Unfortunately, this is the region which shows the largest sensitivity to
 the differences of the parton distribution functions of the photon.
%
%
\subsection{HEAVY QUARK PRODUCTION}
\label{sec:ggheavy}
\begin{figure}[htb]
\begin{center}
{\includegraphics[width=0.65\linewidth]{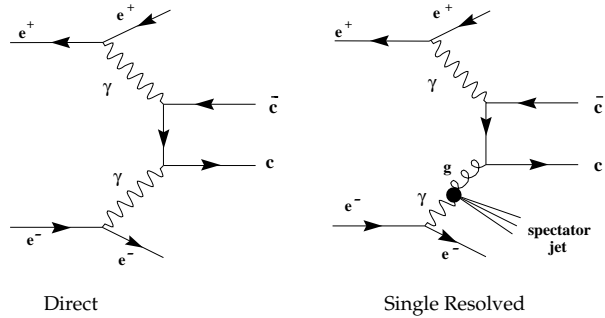}}
\end{center}
\caption{
         The direct (left) and single-resolved (right) contributions to
         charm quark production in photon-photon collisions.
        }\label{fig:fig07}
\end{figure}

 The production of heavy quarks in photon-photon scattering is dominated
 by charm quark production, as the bottom quarks are much heavier and
 have a smaller electric charge.
 Due to the large scale of the process provided by the charm quark mass,
 the production of charm quarks can be predicted in next-to-leading order
 perturbative QCD.
 In QCD the production of charm quarks at LEP2 energies receives 
 contributions of about equal size from the direct production mechanism and
 from the single-resolved contribution, shown in Figure~\ref{fig:fig07}.
 In contrast, the double-resolved contribution is expected to be 
 very small, see Ref.~\cite{FRI-9901} for details.
 \par
%
\begin{figure}[htb]
\begin{center}
{\includegraphics[width=1.0\linewidth]{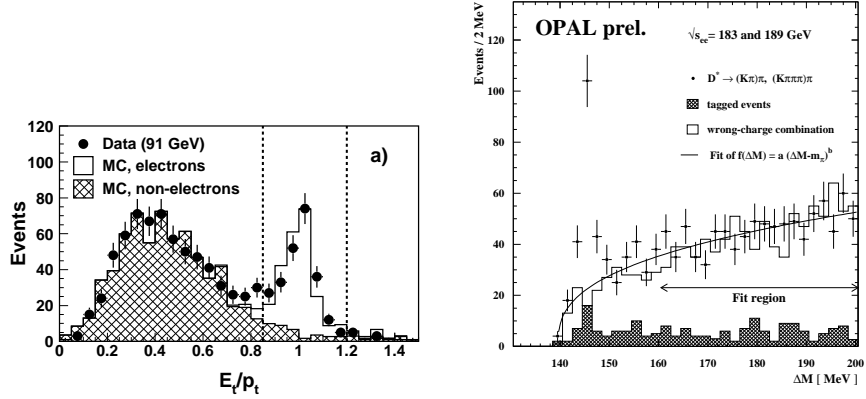}}
\end{center}
\caption{
         Charm quark tagging via electrons from semileptonic decays
         (left), and via the mass difference between the mass of the 
         $D^\star$ and the mass of the $D^0$ candidate (right).
        }\label{fig:fig08}
\end{figure}

%
 In photon-photon collisions the charm quarks have been tagged 
 using standard techniques, either based on the observation of 
 semileptonic decays of charm quarks using identified electrons and muons
 in Ref.~\cite{L3C-9902}, or by the measurement of $D^\star$ 
 production in Refs.~\cite{ALE-9501,PAT-9901,L3C-9905} 
 using the decay $D^\star\rightarrow D^0\pi$, where the pion has
 very low energy, followed by the D$^0$ decay observed in one of the
 decay channels, $D^0\rightarrow K\pi, K\pi\pi^0, K\pi\pi\pi$.
 The leptons as well as the $D^\star$ can be clearly separated from 
 background processes, as shown in Figure~\ref{fig:fig08}.
%
%
 Using the ratio of the transverse energy of the electron measured
 in the calorimeter and the transverse momentum measured in the 
 tracking chamber the electrons can be well separated from other
 charged particles, Figure~\ref{fig:fig08}(left) 
 from Ref.~\cite{L3C-9902}.
 Utilising the low energy of the slow pion a clear peak
 can be observed in the mass difference, $\Delta M$, between the 
 mass of the $D^\star$ and the mass of the $D^0$ candidate, as shown
 in Figure~\ref{fig:fig08}(right) from Ref.~\cite{PAT-9901}.
 However, due to the small branching ratios and selection inefficiencies
 the selected event samples are small and the measurements
 are limited mainly by the statistical error.
 \par
%
\begin{figure}[htb]
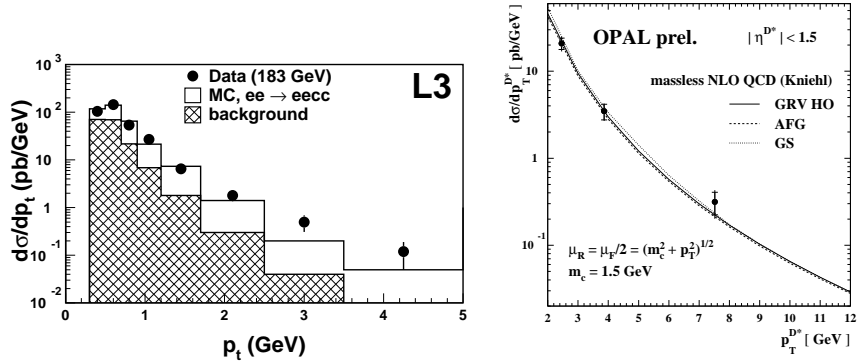

\mbox{}\vspace{-1.1cm}
\begin{center}
{\includegraphics[width=0.52\linewidth]{./nisius.14}}
{\includegraphics[width=0.44\linewidth]{./nisius.15}}
\end{center}
\caption{
         Differential cross-sections for charm quark production with 
         semileptonic
         decays into electrons (left), and for $D^\star$ production (right),
         both determined in restricted kinematical regions.
        }\label{fig:fig09}
\end{figure}
%
 Based on these tagging methods differential cross-sections for charm quark
 production and $D^\star$ production in restricted kinematical regions 
 have been obtained, examples of which are shown in Figure~\ref{fig:fig09}.
 Figure~\ref{fig:fig09}(left) shows the differential cross-section for 
 charm quark production, with semileptonic decays into electrons fulfilling 
 $\vert\cos\theta_{\rm e}\vert<0.9$ and $E_{\rm e}>0.6$~GeV and for
 $W>3$~GeV.
 The data are compared to the leading order prediction from PYTHIA, 
 normalised to the number of data events observed. The shape of the 
 distribution is well reproduced by the leading order prediction.
 Figure~\ref{fig:fig09}(right) shows the differential cross-sections for 
 $D^\star$ production as a function of the transverse momentum of 
 the $D^\star$, for $\vert\eta^{D^\star}\vert<1.5$ compared 
 to the next-to-leading order predictions from Ref.~\cite{BIN-9801}
 calculated in the massless approach.
 The differential cross-sections as functions of the transverse
 momentum and rapidity of the $D^\star$ are well reproduced by the 
 next-to-leading order perturbative QCD predictions, both for the OPAL 
 results~\cite{PAT-9901} and for the L3 results~\cite{L3C-9905}.
 The shape of the OPAL data can be reproduced by the NLO calculations
 from Ref.~\cite{FRI-9901}, however, the theoretical predictions 
 are somewhat lower than the data, especially at low values of
 transverse momentum of the $D^\star$.
 \par
\begin{figure}[htb]
\begin{center}
{\includegraphics[width=1.0\linewidth]{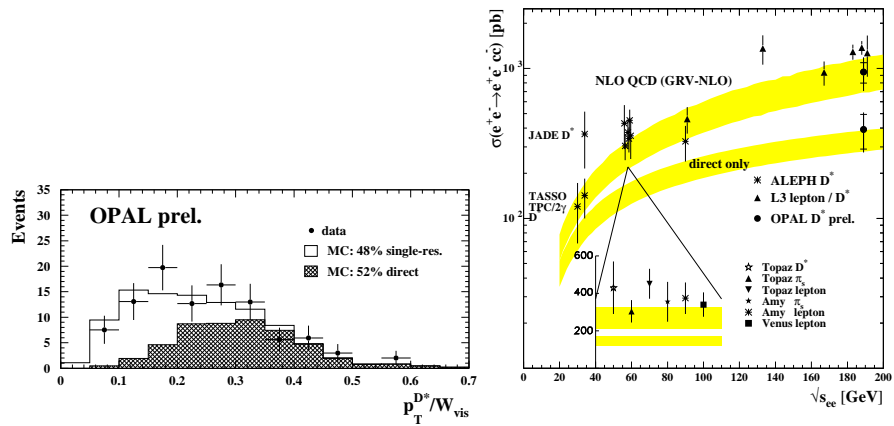}}
\end{center}
\caption{
         The separation of the $D^\star$ production into direct and 
         single-resolved contributions (left) and the total cross-section
         for charm quark production (right).
        }\label{fig:fig10}
\end{figure}

 Based on the observed cross-sections in the restricted ranges in phase
 space the
 total charm quark production cross-section is derived, very much relying
 on the Monte Carlo predictions for the unseen part of the cross-section.
 Two issues are addressed, firstly the relative contribution of
 the direct and single-resolved processes, and secondly the total 
 charm quark production cross-section.
 The direct and single-resolved events, for example, as predicted by
 the PYTHIA Monte Carlo, show a different distribution as a function
 of the transverse momentum of the $D^\star$ meson, $p_T^{D^\star}$, 
 normalised to the visible hadronic invariant mass, $W_{\rm vis}$, as can
 be seen in Figure~\ref{fig:fig10}(left) from Ref.~\cite{PAT-9901}. 
 This feature has been used to experimentally determine the relative
 contribution of direct and single-resolved events, which were found
 to contribute about equally to the cross-section.
 \par
%
%
 The total cross-section for the production of charm quarks is shown in 
 Figure~\ref{fig:fig10}(right). 
 The LEP results are consistent with each other and
 the theoretical predictions are in agreement with the data.
 The measurements suffer from additional errors due to the assumptions  
 made in the extrapolation from the accepted to the total cross-section,
 which are avoided by only measuring cross-sections in restricted
 ranges in phase space.
 It has been shown in Ref.~\cite{FRI-9901}
 that the NLO calculations are flexible enough to account for the
 phase space restrictions of an experimental analysis and that the
 predicted cross-sections in restricted ranges in phase space are less
 sensitive to variations of the charm mass and to alterations of the
 renormalisation as well as the factorisation scale.
 Given this, more insight into several aspects of charm quark production
 may be gained by comparing experimental results and theoretical 
 predictions for cross-sections in restricted ranges in phase space.
 \par
 In addition to the measurements of the charm quark production 
 cross-sections, a preliminary measurement of the cross-section for
 bottom quark production has been reported in Ref.~\cite{NEI-9901}.
%
%
\section{RESULTS FROM $\gamma\gamma^\star$ SCATTERING}
\label{sec:ggs}
 In this kinematical region the reaction can be described as deep-inelastic 
 electron-photon scattering and allows for measurements of photon structure
 functions, similarly to measurements of proton structure functions in
 the case of electron-proton scattering at HERA.
 The measurements of photon structure functions have been discussed in 
 detail in the literature and the reader is referred to the
 most recent review, Ref.~\cite{NIS-9902}, and to references therein.
 Only the main results from the LEP experiments are shortly mentioned here.
 \par
 $\bullet$
 The QED structure function $F_{\mathrm{2,QED}}^{\gamma}$ has been 
 precisely measured using data in the approximate range of average 
 virtualities $\langle Q^2 \rangle$ of $1.5-130$~GeV$^2$. 
 The LEP data are so precise that the effect of the small virtuality
 $P^2$ of the quasi-real photon can clearly be established.
 \par
 $\bullet$
 The structure functions $F_{\mathrm{A,QED}}^{\gamma}$
 and $F_{\mathrm{B,QED}}^{\gamma}$ give more insight into the
 helicity structure of the $\gamma\gamma^\star$ interaction.
 They were obtained from the shape of the distribution of the azimuthal
 angle between the plane defined by the momentum vectors of the muons
 and the plane defined by the momentum vectors of the incoming and the
 deeply inelastically scattered electron.
 Both structure functions were found to be significantly different
 from zero, and the recent theoretical predictions from Ref.~\cite{SEY-9801},
 which take into account the important mass corrections up to 
 ${\mathcal{O}}({m_\mu^2}/{W^2})$, are consistent with the measurements.
 \par
 $\bullet$
 The hadronic structure function $F_{2}^{\gamma}$ has been measured
 using data in the approximate range of average virtualities
 of $\langle Q^2 \rangle$ of $1.9-400$~GeV$^2$. 
 The general features of the measurements can be described by several
 parametrisations of $F_{2}^{\gamma}$.
 However, the data are precise enough to disfavour those parametrisations 
 which predict a fast rise of $F_{2}^{\gamma}$ at low values of $x$, driven 
 by large gluon distribution functions.
 \par
 $\bullet$
 The evolution of $F_{2}^{\gamma}$ with $Q^{2}$ has been studied in bins 
 of $x$. The measurements are consistent with each other and a clear
 rise of $F_{2}^{\gamma}$ with $Q^{2}$ is observed.
 The general trend of the data is followed by the predictions
 of several parametrisations of $F_{2}^{\gamma}$.
 It is an interesting fact that at medium values of $x$ this rise can 
 also be described reasonably well (${\mathcal O}(15\%)$ accuracy) by the 
 leading order augmented asymptotic prediction detailed 
 in Ref.~\cite{OPALPR207},
 which uses the asymptotic solution from Ref.~\cite{WIT-7701}
 for $F_{2}^{\gamma}$ for the light flavour contribution as predicted by
 perturbative QCD for $\alpha_s(M_Z)=0.128$.
 \par
%
%
\section{RESULTS FROM $\gamma^\star\gamma^\star$ SCATTERING}
\label{sec:gsgs}
 The QED and the hadronic structure of virtual photons have been 
 studied at LEP.
 The structure functions of virtual photons can be determined
 for the situation where one photon has a much larger virtuality 
 than the other, $Q^2 \gg P^2$, by measuring the cross-section for 
 events where both electrons are observed.
 For the situation where both photons have similar virtualities,
 $Q^2 \approx P^2$, the structure function picture is no longer applicable
 and differential cross-sections for the exchange of two highly-virtual
 photons have been measured instead.
 The main results from the LEP experiments are shortly mentioned here,
 for a more detailed discussion the reader is referred to 
 Ref.~\cite{NIS-9902}. 
 \par
%
\begin{figure}[htb]
\begin{center}
{\includegraphics[width=1.0\linewidth]{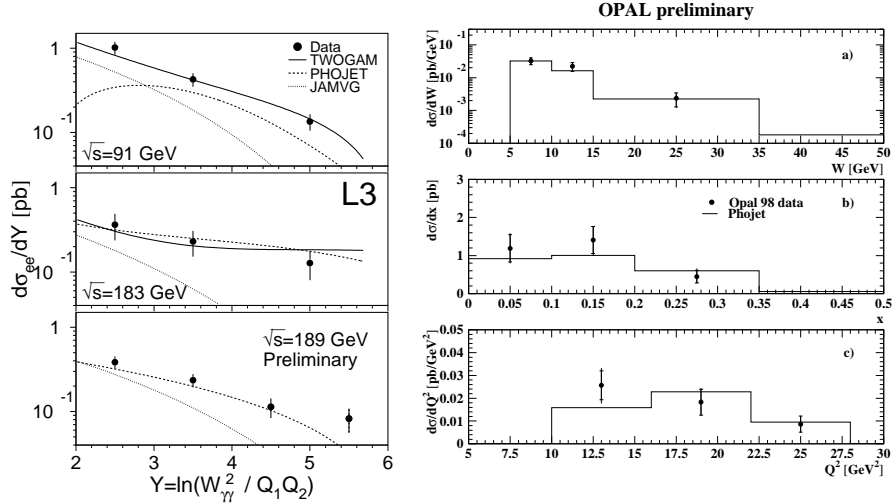}}
\end{center}
\caption{
         The differential cross-sections for the exchange of two highly 
         virtual photons as functions of various variables from
         L3 (left) and OPAL (right).
        }\label{fig:fig11}
\end{figure}
%
 $\bullet$
 The effective hadronic structure function~\cite{PLU-8405} has been measured
 by L3~\cite{ERN-9901} for average virtualities of 
 $\langle Q^2 \rangle = 120$~GeV$^2$ and 
 $\langle P^2 \rangle = 3.7$~GeV$^2$.
 A consistent picture is found for the effective structure function between
 the older PLUTO result from Ref.~\cite{PLU-8405} and the L3 data, and 
 the general features of both measurements are described by the 
 next-to-leading order predictions from Ref.~\cite{GLU-9601}.
 \par
 $\bullet$
 The cross-section for the exchange of two highly virtual photons 
 with muon final states has been measured in Ref.~\cite{OPALPR271}.
 There is good agreement between the measured cross-section
 and the QED prediction.
 The measurement shows that the interference terms, which are 
 usually neglected in investigations of the hadronic structure
 of the photon,
 are present in the data in the kinematical region of the analysis,
 mainly at $x>0.1$, and that the corresponding contributions to the 
 cross-section are negative.
 \par
 $\bullet$
 The cross-section for the exchange of two highly virtual photons with 
 hadronic final states has been
 measured in Refs.~\cite{L3C-9903,ACH-9901,PRY-9901}, and the main 
 results are shown in Figure~\ref{fig:fig11}.
 The differential cross-sections as functions of various variables
 are well described by leading order Monte Carlo models.
 Much larger cross-sections are predicted in the framework of BFKL
 calculations. These predictions are strongly disfavoured by the data.
%
%
 \pagebreak
 \begin{flushleft}{\bf Acknowledgement}\end{flushleft}
 I am grateful to the organisers for inviting me to this inspiring location 
 and for the fruitful atmosphere they created throughout the meeting.
 I wish to thank Stefan S{\"o}ldner-Rembold and Bernd ~Surrow for 
 carefully reading the manuscript and Jochen Patt for providing me 
 with the figure of combined results for the charm quark production 
 cross-section.
%
%
%

%
%
%

\end{document}